# The Nuclear Analogy in AI Governance Research

Sophia Hatz[1]

Department of Peace and Conflict Research/Alva Myrdal Center for Nuclear Disarmament
Uppsala University

**Abstract:** The analogy between Artificial Intelligence (AI) and nuclear weapons is prominent in academic and policy discourse on AI governance. This chapter reviews 43 scholarly works which explicitly draw on the nuclear domain to derive lessons for AI governance. We identify four problem areas where researchers apply nuclear precedents: (1) early development and governance of transformative technologies; (2) international security risks and strategy; (3) international institutions and agreements; and (4) domestic safety regulation. While nuclear-inspired AI proposals are often criticised due to differences across domains, this review clarifies how historical analogies can inform policy development even when technological domains differ substantially. Valuable functions include providing conceptual frameworks for analyzing strategic dynamics, offering cautionary lessons about unsuccessful governance approaches, and expanding policy imagination by legitimizing radical proposals. Given that policymakers already invoke the nuclear analogy, continued critical engagement with these historical precedents remains essential for shaping effective global AI governance.

---

[1] sophia.hatz@pcr.uu.se

# 1. Introduction

Analogies play an important role in shaping our understanding of emerging technologies and their governance. By drawing comparisons to more familiar phenomena, analogical reasoning helps us conceptualize scientific advancements, communicate potential impacts, anticipate future scenarios, and inform the development of policies and institutions (Schwarz-Palschg, 2018). Analogies for Artificial Intelligence (AI) abound, including the invention of fire, first contact with aliens, climate change, the human brain, and nuclear fission (Maas, 2023). In the study of AI governance, AI is often compared to other powerful technologies such nuclear power, the internet, or aviation (Dafoe, 2018, p. 44; Ding, 2024; Ord, 2022; Vermeer, 2024).

The analogy between AI and nuclear technology is particularly prevalent and influential, appearing frequently in academic, policy, and public discourse. Like nuclear technology, AI is dual-use, with the potential to generate vast benefits but also great harm (e.g., Hendrycks et al., 2025; Maas, 2019; Ord, 2022; Zaidi & Dafoe, 2021). Similar to nuclear weapons, the potential risks from advanced AI are societal-scale; next-generation AI models could possess capabilities sufficiently dangerous to jeopardize public safety (Anderljung et al., 2023; Ho et al., 2023). The intense competition for global leadership in AI naturally calls to mind the nuclear arms race between the United States and the Soviet Union during the Cold War (Hendrycks et al., 2025; e.g., Horowitz, 2018; Meacham, 2023).

Consequently, many policy frameworks proposed for AI governance explicitly adapt mechanisms originally developed for nuclear technology, such as an international oversight body similar to the International Atomic Energy Agency (IAEA) (Altman et al., 2023), a government-led effort akin to the Manhattan Project (Tong & Martina, 2024), and an AI research ecosystem modeled on the European Organization for Nuclear Research (CERN) (Rueland, 2025 MAR 17 2025). In contrast, several emerging governance initiatives, including the European Union (EU) AI Act and the Organization for Economic Cooperation and Development (OECD) AI Principles draw on other frameworks, such as risk management from product safety law and principles of responsible innovation. Still, nuclear governance models appear dominant in AI governance, particularly in proposals which address global catastrophic risks and international security concerns.[2]

In academic and policy research, a growing number of studies draw on the nuclear precedent to 'draw lessons' for AI governance. Given the analogy's prevalence and influence, a systematic review of this specific body of academic literature is timely. This chapter provides such a review, focusing on scholarly work which analyzes the nuclear domain—its history, strategic concepts, or governance mechanisms— with the explicit purpose to derive insights for AI governance.

We make two main contributions. First, we provide a comprehensive mapping of how scholars have applied the nuclear precedent across four problem areas in AI governance, synthesizing lessons and identifying patterns in this emerging literature. Second, we adress the broader question of the extent to which the nuclear analogy is valuable, given significant differences across the AI and nuclear domain.

---

[2] For example, note the prevalence in reviews of existing AI governance proposals in Boudreaux et al. (2025), Appendix A and Maas & Villalobos (2023).

Nuclear governance-inspired policy proposals are often criticized on the basis that the domains are too different for the analogy to be useful (Boudreaux et al., 2025; Kaushik, 2023; Klyman & Piliero, 2024; Scharre, 2021; Sepasspour, 2023; e.g., Stewart, 2023). While nuclear weapons are specific, narrow, class of military technology, AI is a General Purpose Technology (GPT) with wide applications across the economy and society (Kaushik, 2023; Klyman & Piliero, 2024; Sepasspour, 2023). Other dissimiliarities include AI's digital and difficult-to-detect nature (Stewart, 2023), the role of non-state actors in research and development (Klyman & Piliero, 2024; Sepasspour, 2023; Stewart, 2023), disagreement about risks (Boudreaux et al., 2025), and its fast and uncertain development trajectory (Sepasspour, 2023).

This review, however, argues that such critiques overlook several key functions of the analogy. While there are many significant differences across AI and the nuclear domain, these differences do not necessarily reduce the value of comparative studies. On the contrary, the central finding of this review is that authors use the AI-nuclear analogy for several functions that do not rely on similarities. The nuclear precedent provides conceptual language and theories that help articulate strategic dynamics—and the different strategic dynamics in each domain—with precision. It offers cautionary lessons into what won't work for AI, given its unique characteristics and context. And it expands policy imagination by legitimizing discussion of ambitious governance proposals. These functions benefit from some commonality between domains but do not require the close similarity that direct policy transfer would demand. In some cases, differences between the domains can enhance analytical value rather than diminish it. For example, discovering that a successful nuclear policy is unlikely to work for AI—due to dissimilarities—is itself an important lesson.

We begin by describing our systematic approach to identifying relevant literature and provide our review. We then examine how scholars conceptualize the AI-nuclear comparison, including similarities and differences across the two domains, and the different analytical functions the analogy serves. We conclude by synthesizing key insights and implications for future research.

## 2. Methods

We conducted a systematic search for literature which explicitly draws on the nuclear domain to derive insights for AI governance. Specifically, we employed a key-word search of Scopus, Web of Science (Core Collection), arXiv, and the web, targeting English-language journal articles, pre-prints and grey literature (reports, white papers and policy briefs).[3] We then scanned the titles, abstracts and introductions of the search results in order to identify articles meeting our inclusion criteria: the primary focus of the article is providing insights into AI risks or governance, and the article explicitly draws on the nuclear domain. In addition to the systematic search, we identified additional articles via the reference lists of selected articles.

In order to provide a review which summarizes the lessons learned from nuclear governance for AI governance, we excluded search results that: a) discuss the implications of the AI-nuclear analogy for AI governance, but do not sufficiently engage with a nuclear case or theory; b) do not explicitly make recommendations for AI governance or AI risk-reduction; or c) argue for existing international agreements or law to expand to include specific AI applications, such as AI-enabled weapons systems or AI in military decision support.

---

[3] Our search strings included keywords referencing the nuclear domain—such "nuclear technology"; "nuclear governance"; "arms control"; "IAEA"; "CERN"; "atoms for peace"; "nonproliferation"–and "artificial intelligence". When searching the web we additionally included terms related to analogical reasoning: "lessons from"; "historical precedent"; "parallels"; "analogy".

Our conceptualisation of the nuclear domain is broad, including nuclear technology, nuclear governance, the history of nuclear technology development and its governance, strategic concepts such as deterrence, Cold War dynamics such as arms races, and specific governance mechanisms such as arms control agreements and the IAEA. While some aspects of the Cold War are not directly linked to nuclear technology, we include Cold War dynamics in the nuclear domain, since we believe the development of nuclear technology and its governance cannot be fully understood outside the Cold War context (see Baum et al. (2022), p. 9; Emery-Xu et al. (2024), p. 14; Backovsky & Bryson (2023), p. 93).

Reflecting the inclusion criteria, the 43 articles we selected for inclusion in the review each include a key phrase in the abstract or introduction linking AI and the nuclear domain, such as "lessons" (Boudreaux et al., 2025; Judge et al., 2025; Law & Ho, 2024; Zaidi & Dafoe, 2021), "parallels" (Bodini, 2024; Maas, 2019; Manheim et al., 2024), "case study" (Baker, 2023; Cha, 2024; Hickey, 2024; Ord, 2022; A. Wasil, 2024), "precedent" (Hendrycks et al., 2025; Trout, 2024), or "modelled on" (Chesterman, 2021).

To structure our review, we categorised the articles according to the governance problem area in focus: (3.1) the early development of transformative technologies and their governance; (3.2) international security risks and strategy; (3.3) international institutions or agreements; and (3.4) domestic approaches to safety regulation.

# 3. Review: what lessons have we learned so far?

## 3.1 Early development and governance of transformative technologies

In this category, researchers examine the uncertain early years of nuclear technology—from the discovery of fission through initial governance attempts—to understand how transformative technologies emerge and how societies first grapple with controlling them (Baum et al., 2022; Boudreaux et al., 2025; Grace, 2015; Ord, 2022; Zaidi & Dafoe, 2021). These studies view AI as an emerging transformative technology that, like nuclear technology before it, promises to fundamentally alter the course of human affairs.

A number of studies examine the period between nuclear fission's discovery (1938) and the atomic bomb's first test (1945) (Grace, 2015; Ord, 2022; Zaidi & Dafoe, 2021). During these years, scientists recognized that an atomic bomb was theoretically possible, spurring various efforts to control or even prevent its development. This historical moment implicitly serves as an analogy for the current AI landscape, where researchers anticipate the imminent arrival of more powerful AI systems, potentially including Artificial General Intelligence (AGI) or Superintelligence (ASI).[4]

The role of secrecy in managing transformative technology emerges as a contested theme. Grace (2015) examines physicist Leó Szilárd's prescient attempts to prevent nuclear catastrophe before the bomb existed, including his efforts to promote secrecy among physicists about fission research and alerting policymakers to the danger (Grace, 2015, pp. 11–15). Grace finds that Szilárd achieved at best limited or ambiguous success; she notes that the secrecy efforts probably contributed to developing further censorship mechanisms and heightened physicists' awareness of their research's applications (Grace, 2015, p. 6). Ord (2022) proves more skeptical, documenting how Soviet spying undermined Manhattan Project security. This failure of secrecy leads Ord to reject proposals that defer AI safety work until just before achieving powerful AI, arguing that even with Los Alamos-level security, key ideas would leak out at a rate similar to the US and UK programmes, and rival teams would remain close behind, leaving minimal time for research on AI safety and control (Ord, 2022, p. 21). Zaidi & Dafoe (2021) further argue that secrecy may play a harmful role in AI policy, arguing it can severely damage organizational deliberation and undermine trust needed for cooperation (Zaidi & Dafoe, 2021, p. 23).

---

[4] AGI generally refers to AI with human-level cognitive abilities and ASI to intelligence surpassing that of humans.

The early nuclear era also demonstrates how technological disruption can temporarily expand political possibilities. Zaidi & Dafoe (2021) characterize the immediate postwar period as a rare historical moment when great powers seriously discussed avoiding an arms race and influential figures genuinely considered relinquishing their monopoly (Zaidi & Dafoe, 2021, p. 6). Between 1944 and 1951, proposals that would normally appear naive or extreme gained serious consideration due to the atomic bomb's spectacular impact and the sense of international crisis (Zaidi & Dafoe, 2021, p. 23). These included Niels Bohr's early appeals to Churchill and Roosevelt, the comprehensive Acheson-Lilienthal Report, and the formal Baruch Plan presented to the United Nations (Zaidi & Dafoe, 2021, pp. 7–8), each of which proposed that nations surrender the technology to an international organization, which would control fissionable raw materials and have a monopoly of all dangerous nuclear activities.

Yet, researchers tend to draw cautionary lessons from the failure of the Baruch Plan. Initial support eroded due to emerging Cold War tensions, sovereignty concerns, excessive secrecy, and domestic political maneuvering (Zaidi & Dafoe, 2021, p. 39). Boudreaux et al. (2025) concludes the Soviet Union rejected the plan because it was perceived as an attempt to cement U.S. nuclear dominance (Boudreaux et al., 2025, pp. 6, 19). Extending lessons to AI, Boudreaux argues that proposals which lock in one nation's technological advantage cannot be expected to look fair or equitable to rivals (Boudreaux et al., 2025, p. 19).

Despite these failures, the historical precedent set by radical proposals serves to expand the boundaries of conceivable AI governance (Zaidi & Dafoe, 2021, p. 2). Emery-Xu et al. (2024) explicitly revive the Acheson-Lilienthal Commission's concept of international control (Emery-Xu et al., 2024). While acknowledging the significant sovereignty implications and lack of modern precedent, they argue that emerging technologies may demand governance solutions of similar scope to those proposed during the early nuclear era (Emery-Xu et al., 2024, p. 3). The fact that radical options were seriously considered contributes a key insight from these historical studies: transformative technologies can create brief windows where ambitious governance proposals become politically viable, even if most ultimately fail (Allen & Chan, 2017, pp. 48–49; Zaidi & Dafoe, 2021, p. 23).

## 3.2 International security risks and strategy

In our second category, the nuclear domain serves as reference for understanding and addressing the international security implications of advanced AI. Scholars and political actors increasingly recognize AI as a dual-use technology capable of conferring significant strategic advantages. This generates concerns that AI could upset power balances, intensify dangerous international competition, and increase catastrophic risks, including great-power conflict (Baum et al., 2022; Black et al., 2024; Hendrycks et al., 2025; Hickey, 2024; Maas, 2019; Stafford et al., 2022). Consequently, literature in this category frames the core governance task as devising national and international strategies that not only manage risks but also fundamentally address actors' strategic interests (Emery-Xu et al., 2024, p. 2; Hendrycks et al., 2025, p. 1). While AI differs from nuclear *weapons* in being a GPT (Hickey, 2024; Maas, 2019), these studies find the nuclear experience a compelling analogy precisely for this strategic challenge, because it provides a historical precedent of a technology forcing radical shifts in strategic thinking under conditions of profound global risk.

With this perspective, several scholars focus on risks and policy in relation to competitive dynamics in AI development, particularly between the U.S. and China. While popular discourse often invokes the US-Soviet nuclear arms race (Imbrie et al., 2020), scholars tend to agree AI competition as whole should not be classified as "arms race", primarily because AI is a GPT with broad applications, not a specific weapon system (Maas, 2019, pp. 286–287).[5] Recognizing this, analyses frequently employ broader terms like 'technological arms race' or 'technology race' to capture the competitive drive surrounding emerging technologies more generally (Stafford et al., 2022, p. 7), or they turn to other relevant international competitions. Stafford et al. (2022) model AI 'technology race' dynamics by combining the security dilemma inherent in nuclear arms races (balancing power projection against catastrophic risk) with features of patent and innovation races (uncertainty, first-mover advantages). This yields new insights for AI, such as safety investments paradoxically increasing risk by encouraging faster deployment of less mature systems, and the particular dangers posed by closely matched competitors. Drawing more broadly on Cold War dynamics, Barnhart (2022) analyzes the US-Soviet Space Race to highlight the role of prestige in driving international technological competition, suggesting potential parallels for US-China AI rivalry (Barnhart, 2022, pp. 42–44).

The nuclear era also serves as a source for strategic foresight, prompting calls for novel thinking analogous to the conceptual shifts demanded by the nuclear revolution (Allen & Chan, 2017, pp. 48–49; Hendrycks et al., 2025, p. 5). Some AI governance proposals adapt specific nuclear strategic concepts, particularly concerning stability between rival powers. Looking towards a potential future with ASI, Hendrycks et al. (2025) adapt the logic of Mutual Assured Destruction (MAD) to propose "Mutual Assured AI Malfunction" (MAIM). Under MAIM, mutual vulnerability to attacks targeting critical AI infrastructure could potentially deter unilateral attempts to achieve ASI dominance. Similarly drawing on deterrence logic, Baum et al. (2022) suggests that credible threats of retaliation for attacks on essential AI infrastructure could enable the relatively peaceful coexistence of rival AI powers (Baum et al., 2022, p. 9).

Beyond specific deterrence concepts, several scholars propose comprehensive national security frameworks for AI. In addition to MAIM, Hendryck et al.'s "Superintelligence Strategy" consists of two additional components: harnessing AI to boost U.S. competitiveness, and preventing the proliferation of dangerous AI to rogue actors. Also focusing on US national security, Allen &

---

[5] See also Grace (2023). At the same time, scholars do agree that an AI technology race could include an arms race in military applications of AI (Geist, 2016; Maas, 2019; Scharre & Lamberth, 2022).

Chan (2017) evaluate past U.S. government paradigms for managing transformative technologies (nuclear, aerospace, cyber and biotech) emphasizing goals such as preserving U.S. leadership, supporting peaceful use, and mitigating catastrophic risks, alongside cross-cutting themes like managing commercial interests (Allen & Chan, 2017, pp. 51–57). Explicitly modeled on Eisenhower's "Atoms for Peace," O'Keefe (2024) proposes a "Chips for Peace" framework built on three commitments: domestic regulation of Frontier AI[6] development to reduce risks, broad sharing of AI benefits, and coordinated restriction of essential inputs to non-participating states (O'Keefe, 2024).

These approaches often recognize that governing AI involves interconnected goals like stability, risk management, and benefit distribution, all requiring some form of technical control (Emery-Xu et al., 2024, p. 2). A key mechanism within several of these frameworks, therefore, involves adapting the logic of non-proliferation and input control strategies (i.e., governing key physical or computational resources essential for development) from the nuclear domain, focusing on chokepoints in AI development. Analogous to the control of fissile materials, attention centers on governing the supply chain for advanced computing hardware (chips) (O'Keefe, 2024; Sastry et al., 2024). Specific proposals detail levers such as compute security, securing model weights, and technical misuse prevention (Hendrycks et al., 2025). The "Chips for Peace" framework (O'Keefe, 2024) operationalizes non-proliferation partly through compute governance and restricting hardware access.

However, these approaches, relying on a "club" of technologically advanced states, inherently risk creating or exacerbating global stratification, a critique also leveled against the nuclear regime (Emery-Xu et al., 2024, p. 15). The political challenges of such models are underscored by the history of nuclear governance. Boudreaux et al. (2025) contrasts the failure of the Baruch Plan—which the Soviet Union rejected as an inequitable attempt to preserve U.S. dominance—with the relative success of the Treaty on the Non-Proliferation of Nuclear Weapons (NPT). The NPT, he argues, only became possible once the superpowers reached a state of mutual vulnerability and found it in their mutual interest to limit further proliferation. A similar basis for multilateral agreement on AI, Boudreaux suggests, remains unlikely until the major powers develop a shared understanding of 'AI stability' and the benefits of collective restraint (Boudreaux et al., 2025, pp. 19, 24).

An alternative, though potentially complementary, strategic perspective emphasizes cooperation, particularly on safety, as a means to manage competitive risks. Ding (2024) uses historical cases of US nuclear safety cooperation (and its failures) with the Soviet Union/Russia and China as analogies to understand the conditions influencing such transfers. The study highlights technological complexity and mutual trust as key factors: for more complex nuclear safety and security technologies, information sharing requires robust technical cooperation and a high degree of trust, as each side risks exposing sensitive information. This offers lessons for potential AI safety collaborations and the feasibility of specific technical proposals like "Permissive Action Links for AI" (Ding, 2024, p. 939).

---

[6] The term "Frontier AI" refers to highly capable foundation models, which could have dangerous capabilities sufficient to pose severe risk to public safety (Anderljung et al., 2023, p. 7).

## 3.3 International institutions and agreements

A third category of literature draws on nuclear governance as a model when proposing international institutions and agreements for managing the global risks and opportunities of advanced AI.[7] Scholars draw inspiration from diverse precedents such as the International Civil Aviation Organization (ICAO), the Intergovernmental Panel on Climate Change (IPCC), the European Organization for Nuclear Research (CERN), international health partnerships, financial standard-setters, and various arms control agreements (Cihon et al., 2020; de Castris & Thomas, 2024; Hausenloy & Dennis, 2023; Ho et al., 2023; Klein & Patrick, 2024; A. R. Wasil, Barnett, et al., 2024). Models derived from the nuclear domain feature prominently, especially when proposals address functions like setting safety standards, monitoring compliance, implementing safeguards, and facilitating verification (Ho et al., 2023; Klein & Patrick, 2024; Nindler, 2019; A. R. Wasil, Barnett, et al., 2024). The IAEA, in particular, is frequently discussed as a model (Ahemd & Kirchschlaeger, 2024; Backovsky & Bryson, 2023; Cha, 2024; Chesterman, 2021; de Castris & Thomas, 2024; Ho et al., 2023; Klein & Patrick, 2024; Law & Ho, 2024; Nindler, 2019; A. Wasil, 2024; see also Maas & Villalobos, 2023, p. 27).

Despite its prevalence, the "IAEA for AI" concept faces significant scrutiny within the literature.[8] While proponents highlight potential functions derived from the IAEA's experience—such as promoting safe use, establishing standards, conducting technical research, performing safeguards, and facilitating cooperation (Ahemd & Kirchschlaeger, 2024; Backovsky & Bryson, 2023; Cha, 2024; de Castris & Thomas, 2024; Vermeer, 2024)—most analyses also stress the model's limitations and the profound differences between the nuclear and AI domains. Common critiques emphasize the IAEA's own historical enforcement challenges (A. Wasil, 2024), the difficulty of monitoring and controlling diffuse, rapidly evolving software compared to tangible nuclear materials (Cha, 2024, p. 12; Chesterman, 2021, p. 202; Law & Ho, 2024, p. 707; Sastry et al., 2024, p. 75), the speed of AI development outstripping institutional response (Hausenloy & Dennis, 2023, pp. 21–22), the prominent role of non-state actors in AI (Hausenloy & Dennis, 2023, p. 22; Zaidi & Dafoe, 2021, p. 5), and the current lack of international consensus on AI risks in contrast to the unified response to the 1945 atomic bombing of Hiroshima (Bodini, 2024, p. 8; Chesterman, 2021, p. 202; A. Wasil, 2024, p. 5).

Given these challenges, many argue that directly replicating any single existing institution is inappropriate (Klein & Patrick, 2024, p. 21; Law & Ho, 2024, p. 707). Instead, AI governance may emerge as a more fragmented "regime complex"[9] involving multiple overlapping institutions and initiatives (Cihon et al. (2020); Klein & Patrick (2024); Geith, this volume). Reflecting this, several proposals advocate for selectively combining functions and lessons from various precedents—including the IAEA, IPCC, CERN, ICAO, and others—to create hybrid or polycentric governance structures tailored to AI's specific challenges (de Castris & Thomas, 2024; Hausenloy & Dennis, 2023; Ho et al., 2023; A. R. Wasil, Barnett, et al., 2024).

---

[7] Maas & Villalobos (2023) provide a literature review of articles proposing new international AI institutions.

[8] There have also been strong criticisms of the IAEA model in popular outlets such as *Foreign Affairs, Foreign Policy, the Wall Street Journal, Vox* and *the Bulletin of Atom Scientists* (Kaushik, 2023; e.g., Stewart, 2023).

[9] A regime complex is generally conceptualised as an array of partially overlapping and non-hierarchical institutions governing a particular issue area.

Researchers also explore potential arms control agreements for military AI, drawing on the history of arms control treaties (e.g., the Intermediate-Range Nuclear Forces (INF) Treaty, the Strategic Arms Reduction Treaty (START), the Chemical Weapons Convention (CWC), and the Biological Weapons Convention (BWC)) (Garcia, 2018; Hickey, 2024; Maas, 2019; Scharre & Lamberth, 2022; A. R. Wasil, Barnett, et al., 2024). While scholars tend to agree competitive AI development should not be characterised as an 'arms race', these studies do call attention to potential arms race dynamics surrounding military applications of AI (Black et al., 2024; Garcia, 2018; Maas, 2019; Scharre & Lamberth, 2022). Recognizing the difficulty of broad bans or verifying software-based systems (Geist, 2016; Maas, 2019; Scharre & Lamberth, 2022), AI arms control proposals often focus on limiting specific, well-defined capabilities based on criteria of mutual desirability and technical feasibility, including verifiability (Scharre & Lamberth, 2022, pp. 4, 12). Where formal verification proves intractable, lessons are drawn from nuclear arms control experiences emphasizing norms, transparency and confidence-building measures (TCBMs), and the role of expert communities (Black et al., 2024; Geist, 2016; Maas, 2019).

The challenge of verification is a dominant theme cutting across institutional and agreement proposals (Emery-Xu et al., 2024, p. 23). Several studies underscore the inherent difficulties posed by AI's characteristics: its software basis, rapid evolution, opacity, dual-use nature, and diffuse development ecosystem (Emery-Xu et al., 2024, p. 14; Geist, 2016, p. 320; Hausenloy & Dennis, 2023, p. 21; Hickey, 2024, pp. 19–20; Maas, 2019, pp. 289–290; O'Keefe, 2024, p. 5; Scharre & Lamberth, 2022, p. 11). While some express skepticism about achieving robust verification comparable to nuclear regimes, Baker (2023) offers a counterpoint. Analyzing nuclear arms control verification successes and failures, Baker argues that verification challenges for AI might be manageable—reduced to levels successfully handled in the nuclear context—if efforts focus on verifiable aspects like hardware inputs (chips) and if appropriate technical methods and institutional precedents are proactively developed (Baker, 2023, pp. 1–3). Similarly, drawing lessons from the INF Treaty, Hickey (2024) argues that enhancing the technical "distinguishability" between civilian and military, or permitted and restricted, AI systems is crucial for enabling future verification regimes (Hickey, 2024, pp. 18–19).

Given the hurdles in verifying AI software and applications, governing key physical inputs—particularly advanced computing hardware—emerges as another significant strategy, analogous to controlling fissile materials in nuclear non-proliferation (Sastry et al., 2024, Appendix A). Sastry et al. (2024) elaborates this 'compute governance' approach, arguing that hardware's tangibility and concentrated supply chain offer governance leverage absent for intangible inputs like data or algorithms (Sastry et al., 2024, p. 2). This focus on input control as an enabler for international governance underpins several frameworks and proposals (Baker, 2023; Emery-Xu et al., 2024; Hendrycks et al., 2025; O'Keefe, 2024). Noting that the compute-Uranium analogy has limits —detecting non-radioactive chips is harder, and the unique proliferation risk from releasing trained model weights lacks a direct nuclear parallel, potentially undermining hardware-centric controls— Sastry et al. primarily focus on adapting the *logic* of input control from the nuclear domain (Sastry et al., 2024, pp. 75–76).

Partly in response to verification difficulties, some propose international agreement structures which are potentially less reliant on highly invasive monitoring. One such model is Emery-Xu et

al.'s 'NPT+', an adaptation of the NPT combining limited proliferation of advanced AI with norms of use and mechanisms for benefit-sharing (Emery-Xu et al., 2024, p. 16).[10] The shift toward hardware-based controls and limited proliferation models reflects a pragmatic recognition that traditional institutions and verification approaches may simply not work for AI, pushing scholars to reimagine fundamental aspects of technology governance.

## 3.4 Domestic safety regulation

In our final category, researchers seek insights from the nuclear domain for ensuring the safety and reliability of AI systems within domestic jurisdictions. Researchers addressing this problem frequently focus on risk management, drawing lessons from the regulation of nuclear power safety and other high-risk sectors such as civil aviation, pharmaceuticals, and finance (Campos et al., 2025; Judge et al., 2025; Manheim et al., 2024; Ortega, 2025; Stein et al., 2024; A. R. Wasil, Clymer, et al., 2024). The core governance challenges motivating these comparisons include establishing effective risk management practices, designing appropriate regulatory oversight structures, assigning liability for potential harms, and creating mechanisms for learning from incidents. This is a relatively new category of articles, perhaps motivated by recent policy initiatives such as the EU AI Act and the establishment of national AI Safety Institutes, which have created a demand for actionable, risk-based policy guidance.

A central lesson emphasized is the need for AI developers and regulatory bodies to proactively prevent harms (Campos et al., 2025, p. 13; Judge et al., 2025, pp. 87–88) by cultivating robust internal risk management processes and a strong organizational safety culture. Elements of being proactive and developing a safety include drawing on the science of AI safety and pre-determining categories of risks as well as acceptable risk thresholds (Campos et al., 2025; Judge et al., 2025, p. 91; Manheim et al., 2024, p. 9; A. R. Wasil, Clymer, et al., 2024, p. 3). Wasil recommends implementing "affirmative safety" principles —a cornerstone of modern nuclear safety regulation— in which developers must demonstrate that their activities keep certain risks below pre-determined thresholds (A. R. Wasil, Clymer, et al., 2024, pp. 1–2). Manheim stresses the need to account for emergent AI capabilities, new misuse opportunities, and evolving technological contexts (Manheim et al., 2024, p. 4) each of which underscore the importance of being proactive.

To tackle the problem of effective government oversight, researchers analyze the structure and powers of agencies like the US Nuclear Regulatory Commission (NRC) and Federal Aviation Administration (FAA) as potential blueprints. Several authors stress the need for a single external oversight body with a clear mandate for auditing or incident response (Judge et al., 2025, p. 91; Manheim et al., 2024, p. 3; Ortega, 2025). Some specific implications for effective oversight include the importance of monitoring across the entire technology lifecycle (including development, testing and deployment), formal verification measures, deep technical expertise within the regulatory body and complementary auditing by private firms (Judge et al., 2025, pp. 89–91; Stein et al., 2024). Recognizing that AI presents unique regulatory challenges (particularly its 'black box' nature, control by tech companies, potentially ambiguous safety goals, and information sensitivity) (Judge et al., 2025, p. 90; Stein et al., 2024), authors tend to recommend adapting the oversight models in their case studies. Adaptations include incorporating formal verification methods where feasible (Judge et al., 2025, p. 92) or using hybrid public-private auditing structures, potentially involving private auditors for governance and security audits under public oversight, while reserving direct public body involvement for safety-critical model evaluations (Manheim et al., 2024; Stein et al., 2024).

---

[10] cf. Klein & Patrick (2024) on the NPT 'peaceful uses' analogy.

Addressing the potential for large-scale harm from AI failures, scholars look to nuclear liability law. The specific regimes developed for the nuclear industry—often involving channeling liability to operators, setting liability limits, and requiring insurance or government indemnity—are examined as precedents for structuring liability for catastrophic AI events (Trout, 2024). Finally, to ensure learning from failures, the incident reporting and investigation systems crucial to safety improvement in nuclear power and aviation are analyzed to inform the design of similar mechanisms for AI accidents and near-misses (Ortega, 2025). Ortega also recommends a proactive approach, suggesting AI developers determine ahead of time which types of events count as reportable incidents (Ortega, 2025, p. 5). Overall, this category is quite optimistic, suggesting that domestic AI safety regulation can build on proven frameworks while adapting to AI's unique challenges.

## 4. Similarities, dissimilarities and functions

The articles reviewed often explicitly reflect on the similarities that motivate the AI-nuclear comparison, the significant differences that limit its applicability, and the inherent challenges of drawing lessons across distinct historical and technological contexts.

Authors typically justify the comparison by identifying salient similarities across the two domains. Common points include the transformative potential of both technologies, their dual-use nature, their origins in scientific breakthroughs, and their strategic significance (Chesterman, 2021, pp. 182–183; Hendrycks et al., 2025, p. 4; Maas, 2019, p. 288; Ord, 2022, p. 1). Often, the argument rests on the magnitude of potential harm; the nuclear experience is seen as the most relevant historical referent for grappling with technologies that pose risks on a global-catastrophic or existential scale (Hendrycks et al., 2025, p. 4; Ord, 2022; Zaidi & Dafoe, 2021, p. 4).

Simultaneously, researchers consistently highlight significant differences that limit the direct transferability of governance mechanisms. These include: AI's intangible software basis versus nuclear physicality; the unprecedented speed of AI progress; the dominant role of private industry versus early state control (Cha, 2024, p. 12; Vermeer, 2024, p. 22); profound monitoring and verification challenges (Baker, 2023, p. 1; Maas, 2019, pp. 289–290); and the lack of global consensus on AI risks comparable to the post-Hiroshima context (Bodini, 2024, p. 8; Chesterman, 2021, p. 202).

Yet, these differences do not necessarily reduce the value of comparative studies. In the works reviewed here, the AI-nuclear analogy serves several distinct functions that do not require the domains to be identical.

First, the nuclear domain provides well-developed theories of deterrence, stability, and arms control, offering a conceptual language to analyze strategic competition in high-stakes contexts (Baum et al., 2022; Hendrycks et al., 2025). This conceptual language works well even when specific conditions differ, and can help us to be more precise about these differences. For instance, Hendrycks et al.'s adaptation of MAD to "Mutual Assured AI Malfunction" (MAIM) demonstrates how nuclear strategic concepts can be reimagined for AI contexts, while Stafford et al. (2022) 's hybrid model combines nuclear security dilemmas with innovation race dynamics to capture AI's unique competitive features.

Second, the analogy is a source of cautionary lessons about "what not to do" (Boudreaux et al., 2025, p. 7). The Baruch Plan's failure warns against AI governance proposals that lock in technological advantages. Ord (2022) 's analysis of Manhattan Project security breaches cautions against deferring AI safety work based on assumptions about maintaining secrecy. In parallel, discovering that a successful nuclear policy is unlikely to work for AI due to dissimilarities also offers a valuable lesson. The extensive critiques of an "IAEA for AI" illustrate this well: scholars consistently highlight how the IAEA's own enforcement challenges, combined with AI's intangible and rapidly evolving nature, make direct institutional replication inappropriate (Cha, 2024; Chesterman, 2021; Law & Ho, 2024). Similarly, traditional arms control verification methods struggle with AI's software basis and dual-use applications, leading researchers to conclude that governing physical inputs like compute hardware may be more feasible than attempting to monitor AI capabilities directly (Baker, 2023; Sastry et al., 2024).

Third, the history of radical proposals for nuclear control expands the space of conceivable policy options for AI, legitimizing discussion of similarly bold ideas like an international monopoly over key resources (Emery-Xu et al., 2024; Zaidi & Dafoe, 2021, p. 2). The fact that figures like Niels Bohr could seriously propose international control of atomic energy to Churchill and Roosevelt, or that the Acheson-Lilienthal Report could envision unprecedented sovereignty concessions, creates precedent for equally ambitious AI governance proposals such as international monopolies over compute resources or the NPT+ framework for limited AI proliferation.

These functions demonstrate that the value of historical comparison extends beyond finding direct policy templates. While these functions benefit from some minimal commonality between domains, they do not require the kind of close similarity that the direct transfer or application of nuclear governance policies demands. Indeed, for the cautionary function, the differences are precisely what generate insight.

Scholars point to other inherent methodological limitations of reasoning from historical precedent. As Ord (2022) notes, history offers no direct counterfactuals, and complex processes rarely unfold identically (Ord, 2022, p. 1). Zaidi & Dafoe (2021) highlights the n=1 problem and potential biases in the historical record (Zaidi & Dafoe, 2021, p. 5). These challenges underscore the need for analytical humility (Allen & Chan, 2017) and caution against overstating conclusions (Vermeer, 2024).

## 5. Conclusion

This systematic review maps a small but growing literature that leverages the history of nuclear technology and its governance to inform contemporary challenges surrounding advanced AI. Across the 43 studies we identified, several key insights and patterns emerge regarding the use and value of this particular historical comparison.

First, the methodological approaches vary considerably. Some scholars undertake detailed historical analyses of specific nuclear cases, time periods, or treaties (Baker, 2023; Ding, 2024; Grace, 2015; Hickey, 2024; Ord, 2022; Zaidi & Dafoe, 2021). Others employ structured comparisons across multiple additional domains (such as aviation, cyber or biotech), seeking common patterns in the governance of high-risk technologies or international cooperation (Allen & Chan, 2017; Cihon et al., 2020; Ho et al., 2023; Judge et al., 2025; Stein et al., 2024; A. R. Wasil, Barnett, et al., 2024). Still others use the nuclear precedent more broadly to frame strategic arguments or motivate specific policy proposals (Hendrycks et al., 2025; O'Keefe, 2024; Trout, 2024).

The review also highlights the diversity of phenomena within the nuclear domain selected for comparison. The "nuclear analogy" is not monolithic; different studies invoke it variously as a precedent for controlling a transformative technology (Ord, 2022), understanding existential risks (Grace, 2015), regulating safety-critical industries (Manheim et al., 2024; Ortega, 2025), regulating technologies of paramount strategic importance (Baum et al., 2022; Maas, 2019), governing dual-use capabilities with grave security implications (Hendrycks et al., 2025; Hickey, 2024), or managing specific sub-domain risks like nuclear facility cybersecurity (Stein et al., 2024). Moreover, analogies are sometimes drawn not just to static features but to specific historical moments or contexts, such as the 'birth of the nuclear age' with scientists raising early warnings (Grace, 2015; Hendrycks et al., 2025; Zaidi & Dafoe, 2021) or the geopolitical situation surrounding Eisenhower's 'Atoms for Peace' address (Law & Ho, 2024; O'Keefe, 2024). This diversity reflects both the nuclear experience and the nature of AI governance challenges.

A third insight relates to the progression or focus of analogies over time. Early studies in our review period often analyzed the initial discovery and uncertain governance attempts of the 1940s, focusing on proactive mitigation and existential fears (Grace, 2015; Ord, 2022; Zaidi & Dafoe, 2021). Strategic analysis drawing on nuclear precedents also appeared early, though often focused on military AI (Allen & Chan, 2017; Geist, 2016; Maas, 2019). More recent work frequently frames AI as a strategically significant dual-use technology, drawing comparisons to later stages of nuclear deployment, regulation (IAEA, NPT), or arms control treaties (Emery-Xu et al., 2024; Hickey, 2024; Manheim et al., 2024; O'Keefe, 2024; Sastry et al., 2024). This apparent shift arguably reflects AI's rapid development, posing challenges for drawing stable lessons—the most relevant analogy might shift faster than the research cycle.

Our central finding is that comparisons with the nuclear domain serve various functions, beyond the pursuit of a policy or institutional template. Some of these functions include drawing on the nuclear domain to provide conceptual language and theories; to identify what will not work; and as a way to expand the policy imagination. Recognizing these distinct functions provides a counterpoint to popular critiques of the analogy's utility based on dissimilarities.

In conclusion, while the AI-nuclear comparison remains contested, its scholarly utility is clear. Given that policymakers already invoke the analogy, continued critical engagement with relevant historical precedents is essential to guide the debate. This research, drawn from the nuclear domain and placed within the broader context of technology governance history, remains an indispensable component of the agenda for global AI governance.

Dafoe, A. (2018). *AI governance: A research agenda*. Centre for the Governance of AI.

de Castris, A. L., & Thomas, C. (2024). The potential functions of an international institution for AI safety. Insights from adjacent policy areas and recent trends. *arXiv*.

Ding, J. (2024). Keep your enemies safer: Technical cooperation and transferring nuclear safety and security technologies. *European Journal of International Relations*.

Emery-Xu, N., Jordan, R., & Trager, R. (2024). International governance of advancing artificial intelligence. *AI & Society*. https://doi.org/10.1007/s00146-024-02050-7

Garcia, D. (2018). Lethal artificial intelligence and change: The future of international peace and security. *International Studies Review*, *20*(2), 334–341. https://doi.org/10.1093/isr/viy029

Geist, E. M. (2016). It's already too late to stop the AI arms race - We must manage it instead. *Bulletin of the Atomic Scientists*, *72*(5), 318–321. https://doi.org/10.1080/00963402.2016.1216672

Grace, K. (2015). *Leó szilárd and the danger of nuclear weapons: A case study in risk mitigation*. Machine Intelligence Research Institute.

Grace, K. (2023). AI is not an arms race. *Time*.

Hausenloy, J., & Dennis, C. (2023). *Towards a UN role in governing foundation artificial intelligence models*. United Nations University Centre for Policy Research.

Hendrycks, D., Schmidt, E., & Wang, A. (2025). Superintelligence strategy: Expert version. *arXiv*.

Hickey, A. (2024). The GPT dilemma: Foundation models and the shadow of dual-use. *arXiv*.

Ho, L., Barnhart, J., Trager, R., Bengio, Y., Brundage, M., Carnegie, A., Chowdhury, R., Dafoe, A., Hadfield, G., & Snidal, M. L. D. (2023). International institutions for advanced AI. *arXiv*. https://doi.org/10.48550/arXiv.2307.04699

Horowitz, M. C. (2018). Artificial intelligence, international competition, and the balance of power. *Texas National Security Review*, *1*(3), 36–57. https://doi.org/10.15781/T2639KP49

Imbrie, A., Dunham, J., Gelles, R., & Aiken, C. (2020). *Mainframes: A provisional analysis of rhetorical frames in AI*. Center for Security and Emerging Technology (CSET) Issue Brief. https://doi.org/10.51593/20190046

Judge, B., Nitzberg, M., & Russell, S. (2025). When code isn't law: Rethinking regulation for artificial intelligence. *Policy and Society*, *44*(1), 85–97. https://doi.org/10.1093/polsoc/puae020

Kaushik, D. (2023). *A "Manhattan Project" for artificial intelligence risk is a bad idea*.

Klein, E., & Patrick, S. (2024). *Envisioning a global regime complex to govern artificial intelligence*. Carnegie Endowment for International Peace.
15